\PassOptionsToPackage{unicode}{hyperref}
\PassOptionsToPackage{hyphens}{url}
\PassOptionsToPackage{dvipsnames,svgnames,x11names}{xcolor}

\documentclass[10pt,a4paper,onecolumn]{article}

\usepackage[T1]{fontenc}
\usepackage[utf8]{inputenc}
\usepackage{lmodern}

\usepackage{marginnote}

\usepackage[rgb,svgnames]{xcolor}
\usepackage{authblk,etoolbox}
\usepackage{titlesec}
\usepackage{calc}
\usepackage{tikz}
\usepackage[pdfa]{hyperref}
\usepackage{hyperxmp}
\hypersetup{%
    unicode=true,
    pdfapart=3,
    pdfaconformance=B,
    pdftitle={JAXMg: A multi-GPU linear solver in JAX},
    pdfauthor={Jacob Tutt, Roeland Wiersema},
    pdfpublication={Journal of Open Source Software},
    pdfpublisher={Open Journals},
    pdfissn={2475-9066},
    pdfpubtype={journal},
    pdfvolumenum={},
    pdfissuenum={},
    pdfdoi={10.xxxxxx/draft},
    pdfcopyright={Copyright (c) 2026, Jacob Tutt, Roeland Wiersema},
    pdflicenseurl={http://creativecommons.org/licenses/by/4.0/},
    colorlinks=true,
    linkcolor=[rgb]{0.0, 0.5, 1.0},
    citecolor=Blue,
    urlcolor=[rgb]{0.0, 0.5, 1.0},
    breaklinks=true
}
\immediate\pdfobj stream attr{/N 3} file{sRGB.icc}
\pdfcatalog{%
  /OutputIntents [
    <<
      /Type /OutputIntent
      /S /GTS_PDFA1
      /DestOutputProfile \the\pdflastobj\space 0 R
      /OutputConditionIdentifier (sRGB)
      /Info (sRGB)
    >>
  ]
}
\usepackage{caption}
\usepackage{orcidlink}
\usepackage{tcolorbox}
\usepackage{amssymb,amsmath}

\usepackage{seqsplit}
\usepackage{xstring}

\makeatletter
\@ifundefined{KOMAClassName}{
  \IfFileExists{parskip.sty}{%
    \usepackage{parskip}
  }{
    \setlength{\parindent}{0pt}
    \setlength{\parskip}{6pt plus 2pt minus 1pt}}
}{
  \KOMAoptions{parskip=half}}
\makeatother
\usepackage{color}
\usepackage{fancyvrb}

\DefineVerbatimEnvironment{Highlighting}{Verbatim}{commandchars=\\\{\}}
\newenvironment{Shaded}{}{}

\newcommand{\DecValTok}[1]{\textcolor[rgb]{0.25,0.63,0.44}{#1}}

\newcommand{\NormalTok}[1]{#1}
\newcommand{\OperatorTok}[1]{\textcolor[rgb]{0.40,0.40,0.40}{#1}}

\newcommand{\StringTok}[1]{\textcolor[rgb]{0.25,0.44,0.63}{#1}}
\newcommand{\VariableTok}[1]{\textcolor[rgb]{0.10,0.09,0.49}{#1}}

\usepackage{graphicx}
\makeatletter
\newsavebox\pandoc@box
\newcommand*\pandocbounded[1]{
  \sbox\pandoc@box{#1}%
  \Gscale@div\@tempa{\textheight}{\dimexpr\ht\pandoc@box+\dp\pandoc@box\relax}%
  \Gscale@div\@tempb{\linewidth}{\wd\pandoc@box}%
  \ifdim\@tempb\p@<\@tempa\p@\let\@tempa\@tempb\fi
  \ifdim\@tempa\p@<\p@\scalebox{\@tempa}{\usebox\pandoc@box}%
  \else\usebox{\pandoc@box}%
  \fi%
}
\def\fps@figure{htbp}
\makeatother
\NewDocumentCommand\citeproctext{}{}
\NewDocumentCommand\citeproc{mm}{%
  \begingroup\def\citeproctext{#2}\cite{#1}\endgroup}
\makeatletter
 \let\@cite@ofmt\@firstofone
 \def\@biblabel#1{}
 \def\@cite#1#2{{#1\if@tempswa , #2\fi}}
\makeatother
\newlength{\cslhangindent}
\newlength{\csllabelwidth}
\newenvironment{CSLReferences}[2] 
 {\begin{list}{}{%
  \setlength{\itemindent}{0pt}
  \setlength{\leftmargin}{0pt}
  \setlength{\parsep}{0pt}
  \ifodd #1
   \setlength{\leftmargin}{\cslhangindent}
   \setlength{\itemindent}{-1\cslhangindent}
  \fi
  \setlength{\itemsep}{#2\baselineskip}}}
 {\end{list}}
\usepackage{calc}

\ifLuaTeX
\usepackage[bidi=basic,shorthands=off,]{babel}
\else
\usepackage[bidi=default,shorthands=off,]{babel}
\fi
\ifLuaTeX
  \usepackage{selnolig} 
\fi
\providecommand{\tightlist}{%
  \setlength{\itemsep}{0pt}\setlength{\parskip}{0pt}}

\usepackage{float}
\let\origfigure\figure
\let\endorigfigure\endfigure
\renewenvironment{figure}[1][2] {
    \expandafter\origfigure\expandafter[H]
} {
    \endorigfigure
}

\usepackage[top=3.5cm, bottom=3cm, right=1.5cm, left=1.0cm,
            headheight=2.2cm, reversemp, includemp, marginparwidth=4.5cm]{geometry}

\titleformat{\section}
  {\normalfont\sffamily\Large\bfseries}
  {}{0pt}{}
\titleformat{\subsection}
  {\normalfont\sffamily\large\bfseries}
  {}{0pt}{}
\titleformat{\subsubsection}
  {\normalfont\sffamily\bfseries}
  {}{0pt}{}
\titleformat*{\paragraph}
  {\sffamily\normalsize}

\usepackage{fancyhdr}
\makeatletter
\let\ps@plain\ps@fancy
\fancyheadoffset[L]{4.5cm}
\fancyfootoffset[L]{4.5cm}

\definecolor{linky}{rgb}{0.0, 0.5, 1.0}

\newtcolorbox{repobox}
   {colback=red, colframe=red!75!black,
     boxrule=0.5pt, arc=2pt, left=6pt, right=6pt, top=3pt, bottom=3pt}

\newcommand{\ExternalLink}{%
   \tikz[x=1.2ex, y=1.2ex, baseline=-0.05ex]{%
       \begin{scope}[x=1ex, y=1ex]
           \clip (-0.1,-0.1)
               --++ (-0, 1.2)
               --++ (0.6, 0)
               --++ (0, -0.6)
               --++ (0.6, 0)
               --++ (0, -1);
           \path[draw,
               line width = 0.5,
               rounded corners=0.5]
               (0,0) rectangle (1,1);
       \end{scope}
       \path[draw, line width = 0.5] (0.5, 0.5)
           -- (1, 1);
       \path[draw, line width = 0.5] (0.6, 1)
           -- (1, 1) -- (1, 0.6);
       }
   }

\definecolor{c53baa1}{RGB}{83,186,161}
\definecolor{c202826}{RGB}{32,40,38}
\def \rorglobalscale {0.1}
\newcommand{\rorlogo}{%
\begin{tikzpicture}[y=1cm, x=1cm, yscale=\rorglobalscale,xscale=\rorglobalscale, every node/.append style={scale=\rorglobalscale}, inner sep=0pt, outer sep=0pt]
  \begin{scope}[even odd rule,line join=round,miter limit=2.0,shift={(-0.025, 0.0216)}]
    \path[fill=c53baa1,nonzero rule,line join=round,miter limit=2.0] (1.8164, 3.012) -- (1.4954, 2.5204) -- (1.1742, 3.012) -- (1.8164, 3.012) -- cycle;
    \path[fill=c53baa1,nonzero rule,line join=round,miter limit=2.0] (3.1594, 3.012) -- (2.8385, 2.5204) -- (2.5172, 3.012) -- (3.1594, 3.012) -- cycle;
    \path[fill=c53baa1,nonzero rule,line join=round,miter limit=2.0] (1.1742, 0.0669) -- (1.4954, 0.5588) -- (1.8164, 0.0669) -- (1.1742, 0.0669) -- cycle;
    \path[fill=c53baa1,nonzero rule,line join=round,miter limit=2.0] (2.5172, 0.0669) -- (2.8385, 0.5588) -- (3.1594, 0.0669) -- (2.5172, 0.0669) -- cycle;
    \path[fill=c202826,nonzero rule,line join=round,miter limit=2.0] (3.8505, 1.4364).. controls (3.9643, 1.4576) and (4.0508, 1.5081) .. (4.1098, 1.5878).. controls (4.169, 1.6674) and (4.1984, 1.7642) .. (4.1984, 1.8777).. controls (4.1984, 1.9719) and (4.182, 2.0503) .. (4.1495, 2.1132).. controls (4.1169, 2.1762) and (4.0727, 2.2262) .. (4.0174, 2.2635).. controls (3.9621, 2.3006) and (3.8976, 2.3273) .. (3.824, 2.3432).. controls (3.7505, 2.359) and (3.6727, 2.367) .. (3.5909, 2.367) -- (2.9676, 2.367) -- (2.9676, 1.8688).. controls (2.9625, 1.8833) and (2.9572, 1.8976) .. (2.9514, 1.9119).. controls (2.9083, 2.0164) and (2.848, 2.1056) .. (2.7705, 2.1791).. controls (2.6929, 2.2527) and (2.6014, 2.3093) .. (2.495, 2.3487).. controls (2.3889, 2.3881) and (2.2728, 2.408) .. (2.1468, 2.408).. controls (2.0209, 2.408) and (1.905, 2.3881) .. (1.7986, 2.3487).. controls (1.6925, 2.3093) and (1.6007, 2.2527) .. (1.5232, 2.1791).. controls (1.4539, 2.1132) and (1.3983, 2.0346) .. (1.3565, 1.9436).. controls (1.3504, 2.009) and (1.3351, 2.0656) .. (1.3105, 2.1132).. controls (1.2779, 2.1762) and (1.2338, 2.2262) .. (1.1785, 2.2635).. controls (1.1232, 2.3006) and (1.0586, 2.3273) .. (0.985, 2.3432).. controls (0.9115, 2.359) and (0.8337, 2.367) .. (0.7519, 2.367) -- (0.1289, 2.367) -- (0.1289, 0.7562) -- (0.4837, 0.7562) -- (0.4837, 1.4002) -- (0.6588, 1.4002) -- (0.9956, 0.7562) -- (1.4211, 0.7562) -- (1.0118, 1.4364).. controls (1.1255, 1.4576) and (1.2121, 1.5081) .. (1.2711, 1.5878).. controls (1.2737, 1.5915) and (1.2761, 1.5954) .. (1.2787, 1.5991).. controls (1.2782, 1.5867) and (1.2779, 1.5743) .. (1.2779, 1.5616).. controls (1.2779, 1.4327) and (1.2996, 1.3158) .. (1.3428, 1.2113).. controls (1.3859, 1.1068) and (1.4462, 1.0176) .. (1.5237, 0.944).. controls (1.601, 0.8705) and (1.6928, 0.8139) .. (1.7992, 0.7744).. controls (1.9053, 0.735) and (2.0214, 0.7152) .. (2.1474, 0.7152).. controls (2.2733, 0.7152) and (2.3892, 0.735) .. (2.4956, 0.7744).. controls (2.6016, 0.8139) and (2.6935, 0.8705) .. (2.771, 0.944).. controls (2.8482, 1.0176) and (2.9086, 1.1068) .. (2.952, 1.2113).. controls (2.9578, 1.2253) and (2.9631, 1.2398) .. (2.9681, 1.2544) -- (2.9681, 0.7562) -- (3.3229, 0.7562) -- (3.3229, 1.4002) -- (3.4981, 1.4002) -- (3.8349, 0.7562) -- (4.2603, 0.7562) -- (3.8505, 1.4364) -- cycle(0.9628, 1.7777).. controls (0.9438, 1.7534) and (0.92, 1.7357) .. (0.8911, 1.7243).. controls (0.8623, 1.7129) and (0.83, 1.706) .. (0.7945, 1.7039).. controls (0.7588, 1.7015) and (0.7252, 1.7005) .. (0.6932, 1.7005) -- (0.4839, 1.7005) -- (0.4839, 2.0667) -- (0.716, 2.0667).. controls (0.7477, 2.0667) and (0.7805, 2.0643) .. (0.8139, 2.0598).. controls (0.8472, 2.0553) and (0.8768, 2.0466) .. (0.9025, 2.0336).. controls (0.9282, 2.0206) and (0.9496, 2.0021) .. (0.9663, 1.9778).. controls (0.9829, 1.9534) and (0.9914, 1.9209) .. (0.9914, 1.8799).. controls (0.9914, 1.8362) and (0.9819, 1.8021) .. (0.9628, 1.7777) -- cycle(2.6125, 1.3533).. controls (2.5889, 1.2904) and (2.5553, 1.2359) .. (2.5112, 1.1896).. controls (2.4672, 1.1433) and (2.4146, 1.1073) .. (2.3529, 1.0814).. controls (2.2916, 1.0554) and (2.2228, 1.0427) .. (2.1471, 1.0427).. controls (2.0712, 1.0427) and (2.0026, 1.0557) .. (1.9412, 1.0814).. controls (1.8799, 1.107) and (1.8272, 1.1433) .. (1.783, 1.1896).. controls (1.7391, 1.2359) and (1.7052, 1.2904) .. (1.6817, 1.3533).. controls (1.6581, 1.4163) and (1.6465, 1.4856) .. (1.6465, 1.5616).. controls (1.6465, 1.6359) and (1.6581, 1.705) .. (1.6817, 1.7687).. controls (1.7052, 1.8325) and (1.7388, 1.8873) .. (1.783, 1.9336).. controls (1.8269, 1.9799) and (1.8796, 2.0159) .. (1.9412, 2.0418).. controls (2.0026, 2.0675) and (2.0712, 2.0804) .. (2.1471, 2.0804).. controls (2.223, 2.0804) and (2.2916, 2.0675) .. (2.3529, 2.0418).. controls (2.4143, 2.0161) and (2.467, 1.9799) .. (2.5112, 1.9336).. controls (2.5551, 1.8873) and (2.5889, 1.8322) .. (2.6125, 1.7687).. controls (2.636, 1.705) and (2.6477, 1.6359) .. (2.6477, 1.5616).. controls (2.6477, 1.4856) and (2.636, 1.4163) .. (2.6125, 1.3533) -- cycle(3.8015, 1.7777).. controls (3.7825, 1.7534) and (3.7587, 1.7357) .. (3.7298, 1.7243).. controls (3.701, 1.7129) and (3.6687, 1.706) .. (3.6333, 1.7039).. controls (3.5975, 1.7015) and (3.5639, 1.7005) .. (3.5319, 1.7005) -- (3.3226, 1.7005) -- (3.3226, 2.0667) -- (3.5547, 2.0667).. controls (3.5864, 2.0667) and (3.6192, 2.0643) .. (3.6526, 2.0598).. controls (3.6859, 2.0553) and (3.7155, 2.0466) .. (3.7412, 2.0336).. controls (3.7669, 2.0206) and (3.7883, 2.0021) .. (3.805, 1.9778).. controls (3.8216, 1.9534) and (3.8301, 1.9209) .. (3.8301, 1.8799).. controls (3.8301, 1.8362) and (3.8206, 1.8021) .. (3.8015, 1.7777) -- cycle;
  \end{scope}
\end{tikzpicture}
}

\patchcmd{\@maketitle}{center}{flushleft}{}{}
\patchcmd{\@maketitle}{center}{flushleft}{}{}
\patchcmd{\@maketitle}{\LARGE}{\LARGE\sffamily}{}{}
\def\maketitle{{%
  
  \AB@maketitle}}
\renewcommand\AB@affilsepx{ \protect\Affilfont}
\renewcommand\AB@affilnote[1]{{\bfseries #1}\hspace{3pt}}
\renewcommand{\affil}[2][]%
   {\newaffiltrue\let\AB@blk@and\AB@pand
      \if\relax#1\relax\def\AB@note{\AB@thenote}\else\def\AB@note{#1}%
        \setcounter{Maxaffil}{0}\fi
        \begingroup
        \let\href=\href@Orig
        \let\protect\@unexpandable@protect
        \def\thanks{\protect\thanks}\def\footnote{\protect\footnote}%
        \@temptokena=\expandafter{\AB@authors}%
        {\def\\{\protect\\\protect\Affilfont}\xdef\AB@temp{#2}}%
         \xdef\AB@authors{\the\@temptokena\AB@las\AB@au@str
         \protect\\[\affilsep]\protect\Affilfont\AB@temp}%
         \gdef\AB@las{}\gdef\AB@au@str{}%
        {\def\\{, \ignorespaces}\xdef\AB@temp{#2}}%
        \@temptokena=\expandafter{\AB@affillist}%
        \xdef\AB@affillist{\the\@temptokena \AB@affilsep
          \AB@affilnote{\AB@note}\protect\Affilfont\AB@temp}%
      \endgroup
       \let\AB@affilsep\AB@affilsepx
}
\makeatother

\renewcommand\Affilfont{\sffamily\small\mdseries}
\PassOptionsToPackage{usenames,dvipsnames}{color} 
\usepackage{lineno}

\ifLuaTeX
  \usepackage{selnolig}  
\fi

\title{JAXMg: A multi-GPU linear solver in JAX}

\author[1,2%
*%
\ensuremath\mathparagraph]{Jacob Tutt%
  \,\orcidlink{0009-0002-5358-4292}\,%
}
\author[3%
*%
\ensuremath\mathparagraph]{Roeland Wiersema%
  \,\orcidlink{0000-0002-0839-4265}\,%
}

\affil[1]{Cavendish Astrophysics, University of Cambridge, Cambridge CB3
0HE, UK%
  \,\protect\href{https://ror.org/0247acz73}{\protect\rorlogo}\,%
}
\affil[2]{Kavli Institute for Cosmology, University of Cambridge,
Cambridge CB3 0HA, UK%
  \,\protect\href{https://ror.org/00pwqz914}{\protect\rorlogo}\,%
}
\affil[3]{Center for Computational Quantum Physics, Flatiron Institute,
162 Fifth Avenue, New York, NY 10010, USA%
  \,\protect\href{https://ror.org/00sekdz59}{\protect\rorlogo}\,%
}
\affil[$\mathparagraph$]{Corresponding author}
\affil[*]{These authors contributed equally.}
\date{\vspace{-2.5ex}}

\begin{document}
\maketitle

\marginpar{

  \begin{flushleft}
  \sffamily\small

  {\bfseries DOI:} \href{https://doi.org/10.xxxxxx/draft}{\color{linky}{10.xxxxxx/draft}}

  \vspace{2mm}
    {\bfseries Software}
  \begin{itemize}
    \setlength\itemsep{0em}
    \item \href{https://github.com/openjournals}{\color{linky}{Review}} \ExternalLink
    \item \href{https://github.com/flatironinstitute/jaxmg}{\color{linky}{Repository}} \ExternalLink
    \item \href{https://doi.org/10.5281}{\color{linky}{Archive}} \ExternalLink
  \end{itemize}

  \vspace{2mm}
  
    \par\noindent\hrulefill\par

  \vspace{2mm}

  {\bfseries Editor:} \href{https://joss.theoj.org}{Open
Journals} \ExternalLink
   \\
  \vspace{1mm}
    {\bfseries Reviewers:}
  \begin{itemize}
  \setlength\itemsep{0em}
    \item \href{https://github.com/openjournals}{@openjournals}
    \end{itemize}
    \vspace{2mm}
  
    {\bfseries Submitted:} 03 September 2026\\
    {\bfseries Published:} Unpublished

  \vspace{2mm}
  {\bfseries License}\\
  Authors of papers retain copyright and release the work under a Creative Commons Attribution 4.0 International License (\href{https://creativecommons.org/licenses/by/4.0/}{\color{linky}{CC BY 4.0}}).

  \end{flushleft}
}

\section{Summary}\label{summary}

Solving large dense linear systems and eigenvalue problems is a core
requirement in many areas of scientific computing, but scaling these
operations beyond a single GPU remains challenging within modern
programming frameworks. While highly optimized multi-GPU solver
libraries exist, they are typically difficult to integrate into
composable, just-in-time (JIT) compiled Python workflows.

JAXMg provides distributed dense linear algebra for JAX, enabling linear
solves and decompositions for matrices that exceed single-GPU memory
limits. By interfacing JAX with NVIDIA's cuSOLVERMp through an XLA
Foreign Function Interface, JAXMg exposes distributed GPU routines as
JIT-compatible JAX primitives. This design allows scalable linear
algebra to be embedded directly within JAX programs, preserving
composability with JAX transformations and enabling multi-GPU and
multi-node execution in end-to-end scientific workflows.

\section{Statement of need}\label{statement-of-need}

Modern scientific computing increasingly relies on GPUs, which now
provide a large fraction of the available floating-point throughput in
both supercomputers and smaller multi-GPU workstations. At the same
time, dense linear algebra remains a critical building block for many
numerical methods. A number of mature libraries therefore provide
high-performance linear algebra on CPUs and GPUs, including distributed
and accelerator-aware packages such as ScaLAPACK
(\citeproc{ref-blackford1997scalapack}{Blackford et al., 1997}), MAGMA
(\citeproc{ref-abdelfattah2024magma}{Abdelfattah et al., 2024}), and
SLATE (\citeproc{ref-gates2019slate}{Gates et al., 2019}).

In parallel, JAX (\citeproc{ref-jax2018github}{Bradbury et al., 2018})
has become a widely adopted framework for scientific computing because
it combines a simple user experience with JIT compilation and automatic
differentiation. The JAX ecosystem has expanded rapidly, with libraries
for neural networks (\citeproc{ref-flax2020github}{Heek et al., 2024}),
Bayesian inference (\citeproc{ref-cabezas2024blackjax}{Cabezas et al.,
2024}), differential equations (\citeproc{ref-kidger2021on}{Kidger,
2021}), variational Monte Carlo (\citeproc{ref-netket3:2022}{Vicentini
et al., 2022}), and full physics simulation environments
(\citeproc{ref-brax2021github}{Freeman et al., 2021}). These workflows
often require repeatedly solving linear systems or computing eigenvalue
decompositions, either as part of a larger simulation loop or inside
differentiable optimization.

Despite this growth, and the availability of packages such as Lineax for
composable linear solves within JAX (\citeproc{ref-lineax2023}{Rader et
al., 2023}), the ecosystem still lacks distributed dense linear solver
routines that scale across multiple GPUs while remaining usable from
idiomatic JAX programs. This gap makes it challenging to take existing
JAX-based scientific applications to problem sizes that exceed a single
GPU, or to integrate multi-GPU linear algebra into more complex JAX
pipelines. Existing approaches require leaving the JAX execution model,
either by exporting arrays to external MPI-based solvers or by manually
orchestrating GPU kernels outside JAX's JIT. These approaches break
composability and complicate memory management.

JAXMg addresses this need by providing a distributed multi-GPU and
multi-node interface to GPU-accelerated solver backends, enabling
scalable linear solves and eigendecompositions from within JAX.

\section{Software design}\label{software-design}

JAXMg connects JAX to NVIDIA's distributed dense linear algebra library
cuSOLVERMp (\citeproc{ref-cusolver}{Corporation, n.d.}) via an XLA
Foreign Function Interface (FFI) C++/CUDA extension. This design enables
writing complex, JIT-compatible JAX programs while delegating the
computationally intensive components to a compiled backend.

Simply pass JAXMg an ordinary JAX array sharded over a two-dimensional
device mesh. The native backend handles the local memory-layout
conversion, 2D block-cyclic redistribution, distributed solver
execution, and restoration of the result to its original JAX layout.

The current release provides a JIT-compatible interface to four
workflows:

\begin{itemize}
\tightlist
\item
  \texttt{potrs}: Solves \(Ax=b\) for symmetric (Hermitian)
  positive-definite \(A\) using a Cholesky factorization
  (\texttt{cusolverMpPotrf} and \texttt{cusolverMpPotrs}). The same
  factorization can optionally return \(\log\det(A)\).
\item
  \texttt{lu\_solve}: Solves \(Ax=b\) for general nonsingular \(A\)
  using a pivoted LU factorization (\texttt{cusolverMpGetrf} and
  \texttt{cusolverMpGetrs}).
\item
  \texttt{syevd}: Computes the eigenvalues \(\lambda_i\) and
  eigenvectors \(v_i\) of a symmetric (Hermitian) matrix \(A\),
  satisfying \(Av_i=\lambda_i v_i\) (\texttt{cusolverMpSyevd}).
\item
  \texttt{gesvd}: Computes the singular-value decomposition of an
  \(M\times N\) matrix \(A=U\Sigma V^\dagger\), returning the singular
  values and optional left and right singular vectors
  (\texttt{cusolverMpGesvd}).
\end{itemize}

All four routines support the JAX dtypes float32, float64, complex64,
and complex128, with CUDA 12 and CUDA 13 backends available for both
x86\_64 and aarch64 systems.

For example, the \texttt{potrs} routine can be called over a
two-dimensional mesh:

\begin{Shaded}
\begin{Highlighting}[]
\NormalTok{mesh }\OperatorTok{=}\NormalTok{ jax.make\_mesh((jax.process\_count(), }\DecValTok{1}\NormalTok{), (}\StringTok{"pr"}\NormalTok{, }\StringTok{"pc"}\NormalTok{))}
\NormalTok{matrix\_specs }\OperatorTok{=}\NormalTok{ P(}\StringTok{"pr"}\NormalTok{, }\StringTok{"pc"}\NormalTok{)}
\end{Highlighting}
\end{Shaded}

Here, \texttt{A} is an \(N\times N\) positive-definite matrix sharded
over the process rows and columns. The solve input \texttt{b} has shape
\(N \times N_{\mathrm{RHS}}\) and is sharded over the process rows:

\begin{Shaded}
\begin{Highlighting}[]
\NormalTok{A }\OperatorTok{=}\NormalTok{ jax.device\_put(A, NamedSharding(mesh, P(}\StringTok{"pr"}\NormalTok{, }\StringTok{"pc"}\NormalTok{)))}
\NormalTok{b }\OperatorTok{=}\NormalTok{ jax.device\_put(b, NamedSharding(mesh, P(}\StringTok{"pr"}\NormalTok{, }\VariableTok{None}\NormalTok{)))}
\end{Highlighting}
\end{Shaded}

The sharded arrays can then be passed directly to \texttt{potrs}:

\begin{Shaded}
\begin{Highlighting}[]
\NormalTok{out }\OperatorTok{=}\NormalTok{ potrs(A, b, T\_A}\OperatorTok{=}\NormalTok{T\_A, mesh}\OperatorTok{=}\NormalTok{mesh, matrix\_specs}\OperatorTok{=}\NormalTok{P(}\StringTok{"pr"}\NormalTok{, }\StringTok{"pc"}\NormalTok{))}
\end{Highlighting}
\end{Shaded}

The tile size \(T_A\) is user-configurable and controls the trade-off
between memory usage and performance; larger tiles typically improve
throughput once the problem is sufficiently large (see Figure
\ref{fig:tile_sweep}).

\subsection{Memory-efficient data
redistribution}\label{memory-efficient-data-redistribution}

Parallelized linear algebra algorithms require a distributed data layout
to ensure proper load balancing of the available computational power
(\citeproc{ref-dongarra1994}{Dongarra et al., 1994}). For JAXMg, the
central challenge is constructing this layout without reducing the
matrix sizes that can be held in aggregate GPU memory. JAXMg therefore
transforms the donated matrix buffers in place and reuses a single
bounded scratch allocation across all stages. Minimizing memory overhead
alone, however, is not sufficient: the redistribution must also use the
available interconnect bandwidth efficiently. Although arbitrary
permutations can be decomposed into fine-grained cycles, repeated small
transfers introduce synchronization and transfer overheads, leading to
poor utilization of the bandwidth available from modern GPU
interconnects (\citeproc{ref-li2020interconnect}{Li et al., 2020}).
JAXMg addresses both requirements through a three-stage redistribution.
Each stage moves the largest contiguous regions that fit within a shared
scratch allocation and performs independent transfers concurrently
wherever dependencies allow. The size of this allocation is determined
by the tile slabs used in the final 2D block-cyclic stage, described in
Section \ref{sec:block-cyclic}, and is reused throughout. The following
sections describe the three forward redistribution stages used to
prepare the matrix for distributed solver execution; after the solver
completes, these stages are reversed to restore the original JAX layout.

\subsubsection{Local memory-layout
conversion}\label{local-memory-layout-conversion}

The first stage reconciles the physical memory layouts used by JAX and
cuSOLVERMp. JAX stores each local matrix shard in row-major order,
whereas cuSOLVERMp requires column-major local buffers. While XLA can
materialize a column-major FFI input, doing so requires a second
full-sized local matrix allocation, defeating the low-memory design.
JAXMg instead applies a parallel implementation of the rectangular
permutation decomposition introduced by Catanzaro et al.
(\citeproc{ref-catanzaro2014transpose}{2014}) directly to each donated
buffer. The method expresses the layout change as modular column and row
permutations, processed in batches bounded by the shared scratch
allocation. The logical matrix remains unchanged and, because the
conversion is entirely local, this stage runs concurrently on every GPU
without inter-device communication.

\subsubsection{Edge-padding alignment}\label{edge-padding-alignment}

\begin{figure}
\centering
\includegraphics[width=1\linewidth,height=\textheight,keepaspectratio]{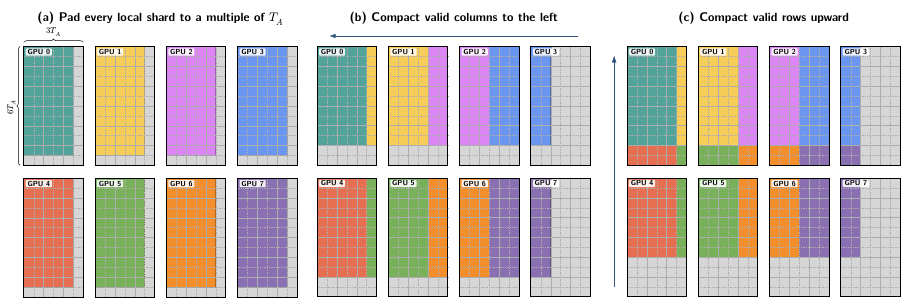}
\caption{Demonstration of the edge-padding alignment across a
\(2 \times 4\) GPU process grid. (a) Initial JAX block-sharded layout
after each local shard is padded to a multiple of \(T_A\). (b)
Horizontal compaction shifts valid column slabs to the left within each
process row. (c) Vertical compaction then shifts valid row slabs upwards
within each process column, leaving all padding on the global right and
bottom edges. Colours indicate the GPU that originally owned each matrix
entry, while grey cells denote padding.\label{fig:padding-alignment}}
\end{figure}

Due to the 2D block-cyclic layout required by the cuSOLVERMp backend,
JAXMg pads a local JAX shard before it enters the native backend if
either dimension is not divisible by the corresponding tile dimension.
This provides enough local capacity for the solver layout, but leaves
padding between neighbouring shards in the global process grid. As a
result, the destination of a solver tile may still contain part of
another tile, so the redistribution in Section \ref{sec:block-cyclic}
cannot yet move complete tile slabs directly. JAXMg therefore compacts
the real matrix towards the global top-left, leaving the padding on the
global right and bottom edges, as illustrated in Figure
\ref{fig:padding-alignment}.

The compaction proceeds in two passes. Column slabs are first shifted
left within each process row, after which row slabs are shifted upwards
within each process column. Since the padding provides empty
destinations, these movements form open chains and do not require an
additional temporary buffer for preserving overwritten data.
Dependencies between movements prevent an entire pass from being
executed at once, so each pass is divided into ordered waves. Within
each wave, the largest slabs that fit in the shared scratch allocation
are moved concurrently across independent process rows or columns.

Unlike the in-place redistribution handled by the native backend, this
initial capacity padding must be performed by JAX because an existing
donated allocation cannot be expanded once assigned. Materializing the
padded array thus temporarily requires both the original and padded
buffers, reducing the matrix size that fits in available GPU memory.
Padding should therefore be avoided where possible by choosing a tile
size that divides both dimensions of every local matrix shard.

\subsubsection{2D block-cyclic redistribution}\label{sec:block-cyclic}

Finally, JAXMg constructs the 2D block-cyclic layout required by
cuSOLVERMp. For a process grid with \(P_r\) rows and \(P_c\) columns,
tiles are distributed in round-robin order over both axes, such that the
tile at global tile coordinate \((i,j)\) is assigned to

\[
\operatorname{owner}(i,j)
=
\left(i \bmod P_r,\;j \bmod P_c\right).
\]

\begin{figure}
\centering
\includegraphics[width=1\linewidth,height=\textheight,keepaspectratio]{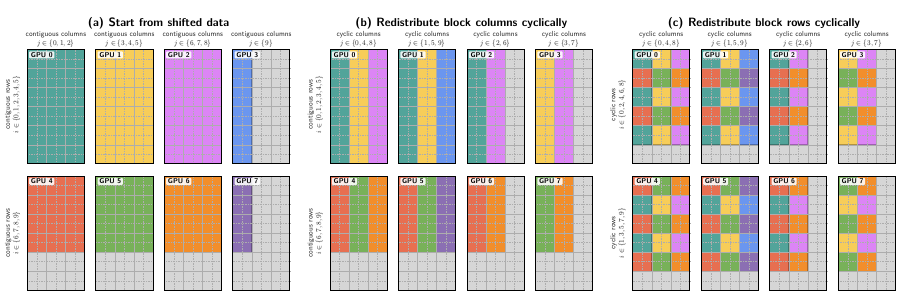}
\caption{Demonstration of the redistribution from the compacted JAX
block-sharded layout to the 2D block-cyclic layout required by
cuSOLVERMp across a \(2 \times 4\) GPU process grid. (a) The
top-left-aligned matrix is partitioned into \(T_A \times T_A\) tiles.
(b) Tile-column slabs are redistributed cyclically within each process
row, assigning global tile column \(j\) to process column
\(j \bmod P_c\). (c) Tile-row slabs are subsequently redistributed
within each process column, assigning global tile row \(i\) to process
row \(i \bmod P_r\). Colours indicate the original GPU ownership of each
matrix entry, while grey cells denote
padding.\label{fig:block-cyclic-redistribution}}
\end{figure}

JAXMg constructs an explicit mapping from every source tile to its
destination and applies it in two separable phases. First, the
column-owner mapping is decomposed into disjoint permutation cycles that
rotate complete tile-column slabs within each process row. The
corresponding cycles run concurrently across all process rows until
every tile has the correct process-column owner. The same procedure is
then applied at the row level within each process column and
parallelized across the process columns, as shown in Figure
\ref{fig:block-cyclic-redistribution}.

During each cycle, a slab is packed into the send scratch slot,
transferred into the receive scratch slot, and unpacked into its
destination, while a third saved slot preserves data that would
otherwise be overwritten before it is forwarded. Let \(N_r\) and \(N_c\)
denote the local row and column capacities, and let \(T_r\) and \(T_c\)
denote the tile dimensions. Since a tile-column slab contains at most
\(T_cN_r\) elements and a tile-row slab contains at most \(T_rN_c\)
elements, the bounded scratch allocation is

\[
S_{\mathrm{scratch}}
=
3\max\left(T_cN_r,\;T_rN_c\right)
\]

elements, where the factor of three accounts for the send, receive, and
saved slots.

\subsubsection{Redistribution
orchestration}\label{redistribution-orchestration}

A central design choice in JAXMg is to build the native backend against
the matching XLA source through Bazel. This gives the FFI handler access
to the underlying NCCL communicator handle (\texttt{ncclComm\_t}) owned
by XLA. JAXMg borrows this communicator for both intra-node and
inter-node redistribution before passing the same handle to cuSOLVERMp
for the distributed solver operation. Movements confined to one rank use
local CUDA operations on the XLA-provided stream. The complete forward
redistribution, solver operation, and reverse redistribution can
therefore share one communication context within a single fused C++/CUDA
FFI call.

\section{Research Impact Statement}\label{research-impact-statement}

\subsection{Scientific applications}\label{scientific-applications}

We first highlight a selection of examples from across physics in which
the dense linear algebra routines provided by JAXMg can be embedded
within wider workflows, allowing scientific applications to scale beyond
the capabilities of a single GPU while remaining within the JAX
ecosystem:

\subsubsection{\texorpdfstring{Cholesky solve
(\texttt{potrs})}{Cholesky solve (potrs)}}\label{cholesky-solve-potrs}

Across statistical inference, positive-definite linear solves are
ubiquitous, particularly in Gaussian processes, a widely used Bayesian
nonparametric model, and in the marginalization of nuisance parameters.
For a Gaussian process with observations \(\mathbf{y}\), mean vector
\(\mathbf{m}\), kernel covariance \(\mathbf{K}\), and noise covariance
\(\boldsymbol{\Sigma}\), the marginal log likelihood is:

\[
  \log \mathcal{L}
  = -\frac{1}{2}(\mathbf{y}-\mathbf{m})^\mathsf{T}
      (\mathbf{K}+\boldsymbol{\Sigma})^{-1}(\mathbf{y}-\mathbf{m})
    -\frac{1}{2}\log\det(\mathbf{K}+\boldsymbol{\Sigma})
    -\frac{N_{\mathrm{data}}}{2}\log(2\pi).
  \]

The quadratic and determinant terms thus require a solve against the
\(N_{\mathrm{data}}\times N_{\mathrm{data}}\) matrix
\(\mathbf{K}+\boldsymbol{\Sigma}\) and its log determinant; analytically
integrating linear nuisance parameters with Gaussian priors yields
analogous operations for their posterior precision matrix. One
application in which both requirements arise is cosmological inference
of the 21-cm brightness-temperature field. As next-generation radio
interferometers such as the Square Kilometre Array (SKA) come online,
JAXMg's distributed Cholesky factorization allows such analyses to
accommodate exascale datasets (\citeproc{ref-liu2026gpr}{Liu et al.,
2026}) and higher-fidelity forward models with larger sets of nuisance
parameters (\citeproc{ref-burba2023allsky}{Burba et al., 2023}) while
remaining within JAX.

Positive-definite solves also arise in stochastic reconfiguration (SR),
a natural-gradient method widely used to optimize variational quantum
states in variational Monte Carlo
(\citeproc{ref-sorella1998green}{Sorella, 1998}). For a state
\(\psi_{\boldsymbol{\theta}}\), SR determines the parameter update
\(\delta\boldsymbol{\theta}\) from

\[
(\mathbf{S}+\lambda\mathbf{I})\delta\boldsymbol{\theta}
= \mathbf{F},
\]

where \(\mathbf{S}\) is the quantum geometric tensor, \(\mathbf{F}\) is
the variational force associated with the energy gradient, and
\(\lambda>0\) is a diagonal regularization. The Monte Carlo estimate of
\(\mathbf{S}\) is Hermitian positive semidefinite; the regularization
makes it positive definite and therefore suitable for Cholesky
factorization. Because this solve is repeated at every optimization
step, explicitly forming \(\mathbf{S}\) for \(N_{\mathrm{par}}\)
parameters requires an \(N_{\mathrm{par}}\times N_{\mathrm{par}}\) dense
matrix and can quickly become limited by single-GPU memory.
Alternatively, in the MinSR formulation
(\citeproc{ref-chen2024empowering}{Chen \& Heyl, 2024};
\citeproc{ref-rende2024simple}{Rende et al., 2024}), the neural tangent
kernel is of size \(N_{\mathrm{samples}}\times N_{\mathrm{samples}}\),
which faces a similar problem as the number of Monte Carlo samples
increases. JAXMg's distributed \texttt{potrs} allows the parameter
update to remain within one JIT-compiled JAX workflow, enabling direct
stochastic-reconfiguration solves for larger neural quantum states and
other parameterized variational ansätze. JAXMg is integrated in the
latest version of NetKet (\citeproc{ref-netket3:2022}{Vicentini et al.,
2022}).

\subsubsection{\texorpdfstring{General linear solve
(\texttt{lu\_solve})}{General linear solve (lu\_solve)}}\label{general-linear-solve-lu_solve}

Across computational electromagnetics, general linear solves are central
to antenna simulations that model electromagnetic fields and their
interaction with complex structures through numerical solutions of
Maxwell's equations. A common approach is the method of moments
(\citeproc{ref-Harrington1993}{Harrington, 1993}), in which the
governing integral equations are discretized to produce the dense
complex system \(ZI=V\), where \(Z\) is the impedance matrix, \(I\)
contains the unknown current coefficients, and \(V\) represents one or
more excitations. Recovering the current coefficients therefore requires
solving this general nonsingular system. One application in which this
becomes computationally demanding is the full-wave simulation of mutual
coupling across the 320-element core of the Hydrogen Epoch of
Reionization Array for 21-cm cosmology
(\citeproc{ref-gueuning2026mutual}{Gueuning et al., 2026}). As precision
requirements drive larger arrays and finer geometric meshes, the number
of basis functions and hence the dimensions of \(Z\) grow. JAXMg's
distributed LU factorization allows the construction of \(Z\), the
solution of \(ZI=V\), and downstream analysis to remain within a single
compiled JAX workflow, supporting efficient end-to-end simulation at
increasing model fidelity.

\subsubsection{\texorpdfstring{Singular-value decomposition
(\texttt{gesvd})}{Singular-value decomposition (gesvd)}}\label{singular-value-decomposition-gesvd}

Principal component analysis (PCA) uses the singular-value decomposition
to identify dominant directions of variation in high-dimensional data.
For a centered data matrix
\(X\in\mathbb{R}^{N_{\mathrm{samples}}\times N_{\mathrm{features}}}\),

\[
X=U\Sigma V^\mathsf{T},
\]

the columns of \(V\) are the principal directions, while \(U\Sigma\)
contains the corresponding low-dimensional representations of the
samples. The variance associated with the \(i\)th component is
proportional to \(\sigma_i^2\), allowing the data to be compressed by
retaining only the leading singular values and vectors. JAXMg's
distributed \texttt{gesvd} routine enables PCA for dense datasets whose
sample or feature dimensions make the full data matrix and its
decomposition too large for a single GPU.

Singular-value decompositions are a central primitive in tensor-network
algorithms for quantum many-body systems
(\citeproc{ref-banuls2023tensor}{Bañuls, 2023};
\citeproc{ref-schollwock2011density}{Schollwöck, 2011}). After applying
a local gate, contracting neighbouring tensors, or enlarging an
auxiliary bond, the resulting tensor is reshaped into a matrix and
decomposed as \(X=U\Sigma V^\mathsf{T}\). Retaining the largest singular
values yields the optimal low-rank approximation in the Frobenius norm,
controls the tensor-network bond dimension, and provides an estimate of
the truncation error through the discarded singular-value weight. The
SVD is therefore the backbone of modern tensor-network algorithms such
as the density matrix renormalization group (DMRG), time-evolving block
decimation (TEBD), and approximate contraction schemes for
high-dimensional networks. As the physical dimension and bond dimensions
grow, the intermediate matrices entering these decompositions can become
considerably larger than the tensors retained after truncation and may
exceed the memory of one GPU. JAXMg's distributed \texttt{gesvd} allows
tensor contractions, reshaping, singular-vector construction, and
truncation to remain within a JIT-compiled JAX workflow, supporting
tensor-network calculations at larger bond dimensions without exporting
arrays to an external distributed solver.

\subsubsection{\texorpdfstring{Symmetric eigendecomposition
(\texttt{syevd})}{Symmetric eigendecomposition (syevd)}}\label{symmetric-eigendecomposition-syevd}

The time-dependent variational principle (TDVP) projects Schrödinger
evolution onto the tangent space of a parameterized quantum state,
producing equations of motion of the form
(\citeproc{ref-Carleo2017}{Carleo \& Troyer, 2017};
\citeproc{ref-Schmitt2020QuantumDynamics}{Schmitt \& Heyl, 2020})

\[
\dot{\boldsymbol{\theta}}
= -i\mathbf{S}^+ \mathbf{F},
\]

where \(\mathbf{S}\) and \(\mathbf{F}\) are the same objects as in the
Cholesky-solve section. In practical time-dependent variational Monte
Carlo calculations, redundant parameters, gauge directions, and Monte
Carlo noise can make \(\mathbf{S}\) singular or severely
ill-conditioned. An eigendecomposition
\(\mathbf{S}=\mathbf{V}\boldsymbol{\Lambda}\mathbf{V}^{\dagger}\)
exposes these directions and permits stable spectral regularization, for
example by discarding eigenvalues below a threshold or replacing
\(\boldsymbol{\Lambda}^{-1}\) with a smooth pseudoinverse. Since the
decomposition may be required at every stage of an ODE integrator, its
memory and computational cost become substantial as the number of
variational parameters grows. JAXMg's distributed \texttt{syevd} enables
full-spectrum regularization and diagnostics to be carried out on
matrices exceeding single-GPU memory while keeping state evaluation,
Monte Carlo estimation, and time integration inside a compiled JAX
program. JAXMg has been used for the t-VMC simulations of
Refs.(\citeproc{ref-Wan2026BlurredSampling}{Wan et al., 2026};
\citeproc{ref-Wiersema2026}{Wiersema, 2026}).

\subsection{Performance and scaling}\label{performance-and-scaling}

\begin{figure}
\centering
\includegraphics[width=1\linewidth,height=\textheight,keepaspectratio]{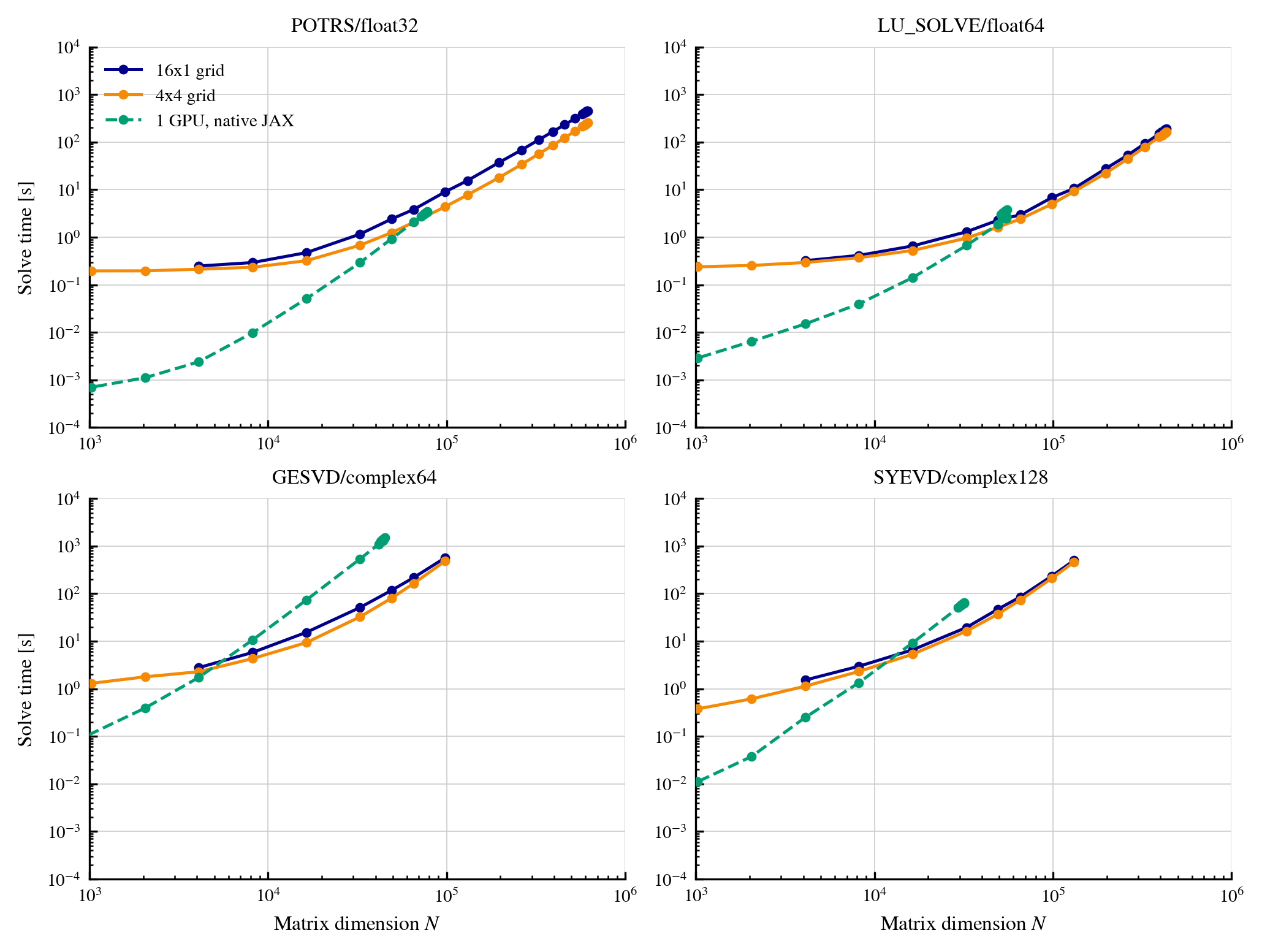}
\caption{Wall-clock runtime against matrix dimension \(N\) for JAXMg on
16 NVIDIA H100 GPUs at fixed tile dimension \(T_A=256\). The
\(16\times1\) process grid is shown in dark blue, the \(4\times4\)
process grid in orange, and native single-GPU JAX in dashed green. The
panels show (a) \texttt{jaxmg.potrs} (\texttt{float32}), (b)
\texttt{jaxmg.lu\_solve} (\texttt{float64}), (c) \texttt{jaxmg.gesvd}
(\texttt{complex64}), and (d) \texttt{jaxmg.syevd}
(\texttt{complex128}). The \(16\times1\) curves begin at larger \(N\)
because avoiding padding requires \(N\) to be divisible by \(16T_A\),
compared with \(4T_A\) for the \(4\times4\) grid.\label{fig:benchmark}}
\end{figure}

\begin{figure}
\centering
\includegraphics[width=1\linewidth,height=\textheight,keepaspectratio]{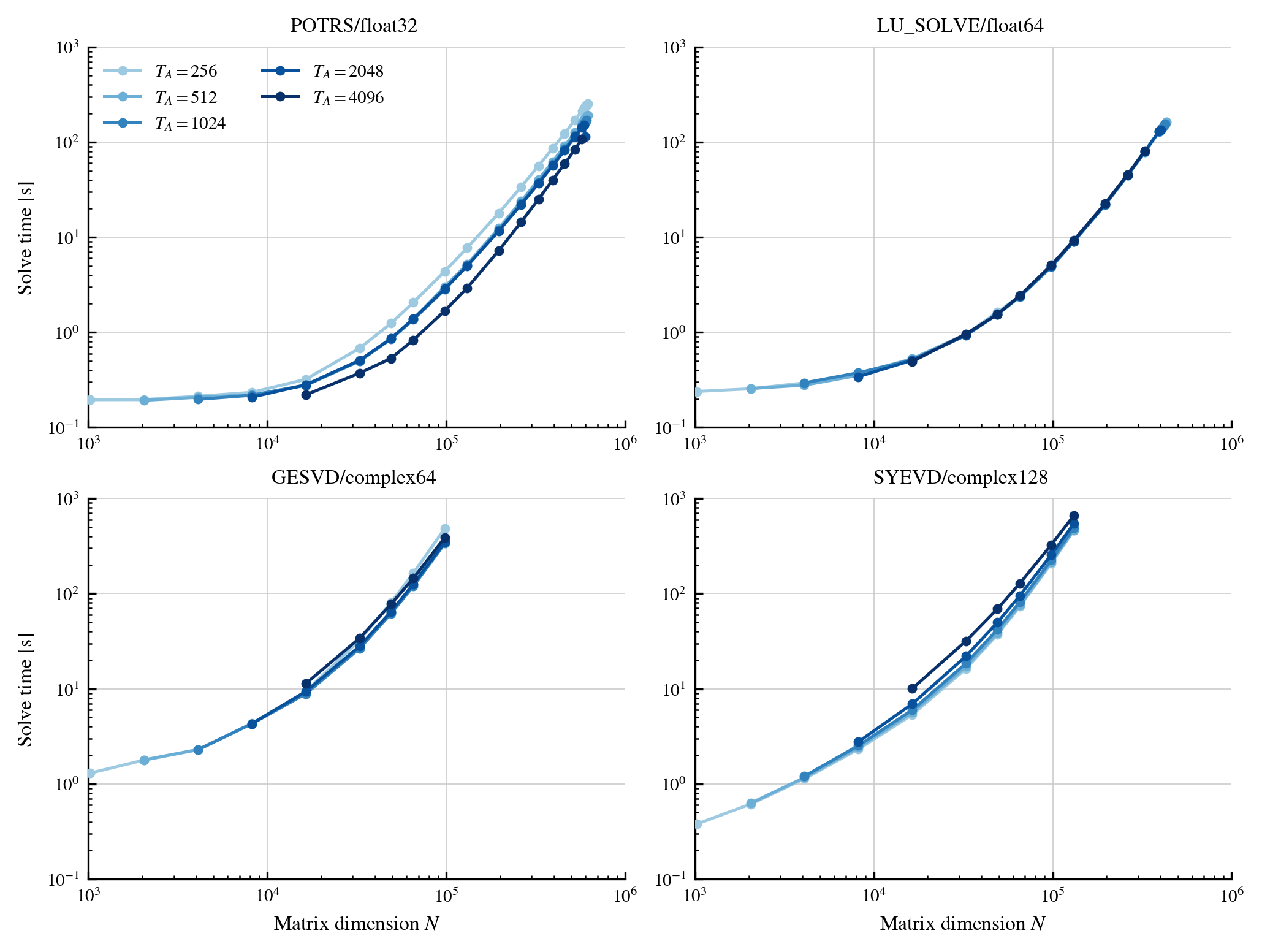}
\caption{Effect of the tile size on JAXMg wall-clock runtime for the
\(4\times4\) process grid on the same 16 NVIDIA H100 GPUs. Each panel
draws one curve per \(T_A\in\{256,512,1024,2048,4096\}\), shaded light
to dark with increasing tile size. There are four panels: (a)
\texttt{jaxmg.potrs} (float32), (b) \texttt{jaxmg.lu\_solve} (float64),
(c) \texttt{jaxmg.gesvd} (complex64) and (d) \texttt{jaxmg.syevd}
(complex128). Curves for larger \(T_A\) begin at larger \(N\) because a
dimension must be a whole number of tiles along both grid
axes.\label{fig:tile_sweep}}
\end{figure}

Performance in distributed matrix routines depends not only on the
available GPU resources, but also on how the calculation is divided
between them. To quantify this dependence and provide practical guidance
for users, we investigate the effects of the process-grid layout and
matrix tile dimension, \(T_A\), on wall-clock runtime in Figures
\ref{fig:benchmark} and \ref{fig:tile_sweep}, respectively.

We consider four representative combinations of routine and data type:
\texttt{potrs} with \texttt{float32}, \texttt{lu\_solve} with
\texttt{float64}, \texttt{gesvd} with \texttt{complex64}, and
\texttt{syevd} with \texttt{complex128}. For \texttt{potrs},
\texttt{lu\_solve}, and \texttt{syevd}, we use the diagonal matrix
\(A=\operatorname{diag}(1,\ldots,N)\); for the two linear solves, we
additionally set \(b=(1,\ldots,1)^\mathsf{T}\). For \texttt{gesvd}, we
use a random Gaussian matrix with zero mean and unit variance. Both
benchmarks use 16 NVIDIA H100 SXM5 GPUs (94 GB each) across four nodes,
with four NVLink-connected GPUs and four 400 Gb/s InfiniBand links per
node. The benchmark implementation is available at
(\citeproc{ref-jaxmg_benchmark}{Tutt \& Wiersema, 2026}).

Figure \ref{fig:benchmark} compares a one-dimensional \(16\times1\)
process grid with a two-dimensional \(4\times4\) grid using a fixed tile
dimension \(T_A=256\). Under the block-cyclic redistribution outlined in
Section \ref{sec:block-cyclic}, moving from the \(16\times1\) to the
\(4\times4\) grid introduces a second inter-rank redistribution phase,
but provides a more balanced two-dimensional distribution during
cuSOLVERMp execution. The process-grid layout therefore determines the
balance between redistribution overhead and distributed solver
efficiency. The results show that, across every routine, the solver-side
benefit outweighs the additional communication, with the \(4\times4\)
grid achieving geometric-mean speed-ups over the matrix dimensions
completed by both layouts of \(1.77\times\) for \texttt{potrs},
\(1.23\times\) for \texttt{lu\_solve}, \(1.38\times\) for
\texttt{gesvd}, and \(1.21\times\) for \texttt{syevd} relative to the
\(16\times1\) grid. The smaller maximum matrix dimensions reached by
\texttt{gesvd} and \texttt{syevd} reflect the significantly larger
cuSOLVERMp workspaces required by these decompositions. Finally, Figure
\ref{fig:benchmark} demonstrates that, for each representative
routine--dtype combination, both distributed layouts outperform the
corresponding native single-GPU JAX implementation at sufficiently large
\(N\), showing that the computational benefit of distributed execution
outweighs its redistribution overhead before single-GPU memory becomes
limiting.

Having established the performance advantage of the \(4\times4\) process
grid, we next hold this layout fixed and vary the tile dimension from
\(T_A=2^8\) to \(2^{12}\), as shown in Figure \ref{fig:tile_sweep}. The
response is strongly routine-dependent: increasing \(T_A\) reduces the
runtime of \texttt{potrs}, leaves \texttt{lu\_solve} broadly unchanged,
and increases the runtime of \texttt{syevd}, while \texttt{gesvd} is
non-monotonic, running fastest at an intermediate \(T_A=1024\) and more
slowly for both smaller and larger tiles.

We finally demonstrate the large-scale performance of JAXMg using 64
NVIDIA H200 SXM5 GPUs (143 GB each) across eight nodes, with eight
NVLink-connected GPUs and eight 400 Gb/s InfiniBand links per node. By
reducing the redistribution scratch required for a fixed tile dimension
relative to the \(64\times1\) layout, the balanced \(8\times8\) process
grid enabled a Cholesky solve with \(N=1{,}499{,}136\) and \(T_A=1024\).
The distributed matrix occupied 8.2 TiB in aggregate, while each local
shard and its redistribution scratch used 94.7\% of the available device
memory. The warm solve completed in 654 seconds. At \(N=1{,}310{,}720\),
the largest dimension completed by both layouts, the \(8\times8\) grid
was also \(6.8\times\) faster, requiring 452 seconds compared with 3065
seconds for \(64\times1\).

These results highlight JAXMg's primary impact: enabling dense matrix
solves and decompositions that are bottlenecked by the memory capacity
of a single GPU, while remaining within JAX's composable and
JIT-compiled programming model. On modern multi-GPU systems, distributed
solvers make it possible to tackle matrix sizes that would otherwise be
infeasible, and to increase throughput by using aggregate device memory
and compute.

\section{AI usage disclosure}\label{ai-usage-disclosure}

Claude Code was used during software development for code exploration
and debugging. Large language models were used to assist with language
polishing.

\section{Acknowledgements}\label{acknowledgements}

We want to thank Dennis Bollweg, Alex Chavin, Geraud Krawezik, Dylan
Simon and Nils Wentzell for their help with developing the code. We also
want to acknowledge the help of Ao Chen and Riccardo Rende with testing
the code in applied settings. RW is grateful to Simon Tartakovsky for
his suggestions on the 1D cyclic algorithm. Finally, we want to thank
Filippo Vincentini for his suggestions on code distribution. RW
acknowledges support from the Flatiron Institute. The Flatiron Institute
is a division of the Simons Foundation. JT is supported by the Harding
Distinguished Postgraduate Scholars Programme (HDPSP) and the Science
and Technology Facilities Council (STFC) DTP Studentship.

The authors acknowledge the use of resources provided by the Isambard-AI
National AI Research Resource (AIRR). Isambard-AI
(\citeproc{ref-Isamabrd_2024}{McIntosh-Smith et al., 2024}) is operated
by the University of Bristol and is funded by the UK Government's
Department for Science, Innovation and Technology (DSIT) via UK Research
and Innovation; and the Science and Technology Facilities Council
{[}ST/AIRR/I-A-I/1023{]}.

\section*{References}\label{references}
\addcontentsline{toc}{section}{References}

\protect\phantomsection\label{refs}

\end{document}